\documentstyle[12pt]{article}
\begin{document}
\setlength{\baselineskip}{2em}
\begin{titlepage}
\begin{center}
{\Large\bf On Bose-Einstein Condensation in Any Dimension} 
\vskip 1.0 cm

{\bf H. Perez Rojas}$^{1}$,\renewcommand{\thefootnote}{*}\footnote
{Permanent address: Grupo 
de Fisica Te\'orica, ICIMAF,
Academia de Ciencias de Cuba, Calle E No. 309,Vedado, La Habana 4, Cuba.}
\vskip 1.5cm
$^1$ {\it International Centre for Theoretical Physics\\
P. O. B. 586, Strada Costiera 11,
34100 Trieste, Italy.}       
\end{center}

\begin{abstract}                                                             
\noindent
Arbitrarily large ground state population is a general property of any ideal 
bose gas
when conditions of degeneracy are satisfied; it occurs at any dimension D.
For $D = 1$, the condensation is diffuse, at $D = 2$ it is a sort of 
quasi-condensate. The discussion is made by following a microscopic 
approach and for finite systems. Some astrophysical consequences are 
discussed, as well as the temperature-dependent mass case.
\end{abstract} 
\noindent 
\end{titlepage}

\newpage

\section{What is Bose-Einstein condensation?}

At present there is a renewed interest in 
Bose-Einstein condensation (BEC), particularly after the experimental 
realization of it \cite{Anderson}. Actually, BEC
is one of the most interesting problems of
quantum statistics. It 
occurs in a free particle Bose gas at a critical temperature $T_c$, and 
is a pure quantum phenomenon, in the sense that no interaction is needed 
to be assumed
to exist  among the particles. BEC is interesting for condensed matter 
(superfluidity, superconductivity) but is has also increasing interest in 
 high energy physics
(electroweak phase transition, superfluidity in neutron stars). The 
consequences of its occurrence at dimensions different from $D = 3$ may have
interest equally in these two fields of physics. 

Bose -Einstein condensation is understood as the steady increase of 
particles in the state with zero energy \cite{Landau}, or as the 
macroscopically large number of particles accumulating in a single 
quantum state \cite{Pathria}, and its connection with the theory of phase 
transitions is actually a property of BEC in  dimensions $D > 2$, since it
has a critical temperature at which the phenomenon of condensation 
starts. 
But as different from BEC, phase 
transitions theory assumes, in general, some interaction among the 
particles  \cite{Pathria}, and properties 
of non-analyticity of thermodynamic quantities appear in the thermodynamic 
limit $N = {\cal N}/V_{{\cal N}, V \to \infty}$, where ${\cal N}$ and $V$
are respectively the number of particles and volume of the system.
It has been also investigated the
possible connection of BEC with spontaneous symmetry breaking (SSB) 
\cite{Haber},
\cite{Kapusta}.
Actually, there is a close analogy, 
but not a full correspondence among them. The SSB assumes also interaction
among the fields, i.e., systems with infinite number of degrees of 
freedom. In systems of low dimensionality, it
happens that no  SSB of a continuous symmetry occurs in one or two 
spatial dimensions $D$ according the the Mermin-Wagner theorem
\cite{Wagner}; (see also \cite{Coleman}, for a proof that there are no 
Goldstone bosons in one dimension). Thre is, however, a close 
correspondence between phase transitions theory and SSB. Concerning BEC, 
it is usually stated \cite{Uhlenbeck} that in the thermodynamic limit
BEC is not possible in  $D = 2$ and  that it neither occurs for $D = 1$.
In the present letter we want to consider again the occurrence of BEC in 
$D = 1,2,3$ and later in any  dimension for the case of systems having
a  finite volume and number of particles.
We will adopt the procedure of investigating 
the microscopic behavior of the density (in momentum space)
which exhibit some interesting 
properties. We also discuss 
an example in the context of astroparticle physics in which there
can be observable phenomena which would escape in the thermodynamic limit.
We must mention finally that the 
general case of condensation in an arbitrary dimension 
was first studied  by May \cite{May} for $D \geq 2$ and later by Ziff, 
Uhlenbeck and Kac \cite{Uhlenbeck}.

The Bose-Einstein distribution $(e^{(E - \mu)/T} - 
1)^{- 1}$ for $\mu < 0$ and $p \neq 0$ vanishes strictly at $T = 0$, which 
fact suggests that 
at $T = 0$ no excited states of a Bose gas can exist on the average  and
{\it condensation in the ground state seems to be a general property, 
whenever the conditions of
quantum degeneracy of the Bose-Einstein gas are satisfied}. Quantum 
degeneracy is 
usually understood to be achieved when the De Broglie thermal wavelength 
$\lambda$ is greater that 
the mean interparticle separation $N^{-1/3}$. However, the remarkable 
discovery made by Einstein on the Bose distribution was that condensation 
may occur at temperatures different from zero, which is what is 
usually properly named BEC. 

According to our previous considerations,
there are two different ideas 
which usually are considered to be the same, concerning what is to be 
understood as BEC: 1) 
The existence of a critical 
temperature $T_c > 0$ such that $\mu(T_c) = E_0$, where $E_0$ is the 
single particle ground state energy. (This condition is usually taken as a 
necessary and 
sufficient condition for condensation; see i. e. \cite{Kirsten}). Then for  
$T \leq T_c$, some 
significant amount of particles starts to condense in the ground state. 
2)The existence a finite fraction of the total 
particle density in the ground state and in states in some neighborhood of 
it at some  temperature 
$T > 0$. We shall name 1) the strong and 2) the weak criterion.

From the point of view of finite-temperature quantum field theory, 
the strong criterion for BEC leads
to the infrared $k^{-2}$ divergence of the 
Boson propagator, 
\[
\lim_{k_4 < {\bf k}\to 0} G (k_4 - i \mu, {\bf k},  T) \simeq 1/-\mu^2 + 
E_0^2 \simeq 1/k^2, 
\]
which is cancelled by the density of states $4 \pi k^2$ when calculating 
the particle density. High temperature radiative corrections 
usually have mainly the
effect of shifting the (longitudinal) boson mass in an amount $\delta M^2 
\simeq T^2$  (Debye screening) and although many 
physical features in the investigation of Bose -Einstein 
condensation would appear in the non-relativistic limit (and in the 
one-loop approximation), we would discuss at the end briefly some features
of the relativistic limit.

\section{The critical temperature in the $D = 3$ case}

Let us remind the origin of the critical quantities 
$\mu_c$, $T_c$ in the standard $3 D$ theory 
of BEC.
The chemical potential $\mu = f(N, T) < 0$ is a decreasing function of 
temperature at 
fixed density $N$, and for $\mu = 0$ one gets an equation defining $T_c 
=f_c (N)$. For temperatures $T < T_c$, as $\mu = 0$, the 
expression for 
the density gives  values $N'(T) < N$, and the difference $N - N'= N_0$ is 
interpreted as the density of particles in the condensate. The mean 
interparticle separation is then $l = N^{-1/3}$. 

In our considerations we will use integrals, as usually, understood as 
approximations of sums over discrete quantum states, without implying
to take the thermodynamic limit. For usual macroscopic systems, as the
separation between quantum states is  $\Delta p = h/V^{1/3}$, the 
approximation of the sum by the integral quite justified.

Now, above the critical temperature for condensation
\begin{eqnarray}
N &=& 4 \pi \lambda^{-3}\int_0^{\infty}\frac{x^2 d x}{e^{x^2 + \bar \mu } 
- 1} \\[1em]
&&\mbox{}= \lambda^{-3} g_{3/2}(z)
\end{eqnarray}
\noindent
where $\bar \mu = - \mu/T ( > 0)$, $x = p / p_T$,  is the relative momentum 
 $p_T = \sqrt{2 m T}$ being the characteristic 
thermal momentum and $\lambda = h / (2 \pi m T)^{1/2}$ is the De Broglie 
thermal wavelength. The function $g_n (z)$ 
is (see i. e. \cite{Pathria})
\[
g_n (z) = \frac{1}{\Gamma(z)} \int_0^\infty \frac{x^{n - 1} d x}{z^{- 1} 
e^x - 1}.
\]
where $z = e^{\mu/T}$ is the fugacity. At $T = T_c$, we have $g_n 
(0) = \zeta (n)$, and  $g_{3/2} (0) = \zeta (3/2)$, and the density 
is 
\begin{equation}
N_c = \frac{\zeta (3/2)}{\lambda^3}
\end{equation}
or in other words, $l^{-3} \lambda^{3} = \zeta(3/2) \simeq 2.612$.
Let us have a microscopic look at  BEC in the $D = 3$ case
and to this end we investigate  in detail the following quantity defined as
the particle density in relative momentum  space 
\[
f_3 (x, \bar \mu) = \frac{x^2}{e^{x^2 + \bar \mu} - 1}
\]
By calculating the first and second derivatives of this function, we 
find that for $\bar \mu \neq 0$it has a minimum for $x = 0$ and a 
maximum for  $x = x_\mu $ where $x_\mu$ is the solution of 
\[
e^{x^2 + \bar \mu} = \frac{1}{(1 - x^2)}.
\] 
In this sense, $f(x, \bar \mu)$ behaves in a very similar form to
the Maxwell-Boltzmann distribution of classical statistics. But
as $\bar \mu \to 0$, $x_\mu \to 0$ also, and
the  maximum of the density, for strictly $\bar \mu = 
0$, is located at $x = 0$. The convergence to the limit $ x = 0$ is not 
uniform. 
A finite fraction of the total density falls in the ground state.

If we go back and substitute the original integral over momentum by a sum 
over shells of quantum states of momentum (energy states), and write
\begin{equation}
N_c = \frac{4 \pi}{h^3} \sum_{i = 0}^{\infty} \frac{p^2_i \Delta 
p}{e^{p_i^2/2 M T} - 1}.
\end{equation}
\noindent
For $i = 0$, by taking $\Delta p \simeq h/V^{1/3}$, where $V$ is the 
volume 
of the vessel containing the gas, the contribution of the ground 
state density is $N_0 = \frac{4}{V^{1/3} \lambda^2}$.
We have thus a fraction of
\begin{equation} 
N_0/N_c = 4 \lambda \zeta (3/2)/V^{1/3} = 4 \zeta (3/2)/{\cal N}^{1/3} 
\label{sen} 
\end{equation}
\noindent
particles in the ground state, which is the most populated, as described by 
the statistical distribution, at $T = T_c$.
A numerical estimation for one liter of $He$ gas leads to $N_0/N \simeq 
10^{-6}$. In quantum states in a small neighborhood of the
ground state, the momentum density has slightly lower values. Thus, at 
the critical temperature for BEC, there is a set of states close to the
ground state, having relative large densities.

The reader may argue that in the thermodynamic limit (\ref{sen}) vanishes.
It is true. But (\ref{sen}) indicates an interesting relation: the larger the
separation between quantum states, the larger the population in the 
ground state at the critical temperature. In the thermodynamic limit the 
quantum states form a continuum, and (\ref{sen}) has no meaning. However,
all systems of physical interest, in laboratory as well as in 
astrophysical and cosmological contexts have finite $V$ and ${\cal N}$.

 We conclude that {\it at the critical 
temperature for condensation 
the density of particles in momentum space reaches its maximum at zero 
momentum, and by describing the density as a sum over quantum states,
a macroscopic fraction is obtained for the density in the ground 
state and in neighbor states}. Thus, at the critical 
temperature, the weak criterion is satisfied.
 For values of $T < T_c$, the curve describing the 
density in momentum 
space flattens on the $p$ (or $x$) axis, and its maximum decreases also. 
As we admit conservation of particles, we get increased the
ground state density by adding  to $f(x)$  the 
quantity $2 N [1 - (T/T_c)^{3/2}] \theta (T_c - T)\lambda^3 \delta (x)$ 
as an additional density. The density in neighbor states decreases as 
well as in larger momentum states. This leads to the usual Bose-Einstein 
condensation. We must remark that in this case for values of $T$ smaller 
but close to $T_c$ both the weak and the strong criteria are satisfied.

\section{The $D = 1$ and $D = 2$ cases}

Let us see what happens for $D = 1$. In this case, the mean interparticle 
separation $l = N^{-1}$. The density in momentum space coincides with the 
Bose- Einstein 
distribution, $ f_1 (p, T, \mu) = (e^{p^2/2M T - \mu/T} - 1)$. This 
function has only one extremum, a maximum, at $p = 0$. By using the
previous change of variables, we have the expression for the density of 
particles as

\begin{equation}
N =  2 \lambda^{-1} \int_0^\infty \frac{d x}{e^{x^2 + \bar \mu} - 1}=
\frac{1}{\lambda} g_{1/2} (z).
\end{equation}

We have thus that $N \lambda \simeq g_{1/2}(z)$. The fact that $\lim 
g_{1/2} (z)$ diverges as $z \to 1$ indicates an enhancement of the 
quantum degeneracy regime. But this fact actually means that
 $\mu$ is a decreasing function of $T$ for $N$ constant,
and for very small $\bar \mu$ one can write, approximately
\begin{equation}
N \simeq 2 \lambda^{- 1} \int_0^{x_0} \frac{d x}{x^2 + \bar \mu} 
\simeq \frac{\pi \bar \mu^{- 1/2}}{ \lambda}, \label{1}
\end{equation}
where $x_0 = p_0/M T$, $p_0$ being some characteristing momentum $p_0 \gg 
p_T$.
Thus, $\bar \mu$ does not vanishes at $T \neq 0$, and for small $T$ it is 
approximately given by $\bar \mu = \pi^2/4 N^2 \lambda^2$. 
It is easy to obtain  approximately the  pressure $P$ (= force$/L^0$) and 
energy density repectively as
\begin{equation}
P \simeq \frac{\pi^2 T}{N \lambda^2} \hspace{1cm} U = \frac{1}{2} P
\end{equation}
which indicates that they vanish as $T/N \to 0$, the specific heat 
$C_v = \partial U/\partial T$ decreasing as $T^{1/2}$ as $T \to 0$. For high 
temperatures one can easily 
prove that $C_v$ is constant; it indicates some correspondence with the 
behavior of $C_v$ in the $D = 3$ case \cite{Pathria}, but there is no 
discontinuity in its derivative with regard $T$.

By substituting 
the last expression for $\bar \mu$ back  in (\ref{1}), one has that 
\begin{equation}
N \simeq \frac{\lambda^{-1}}{\gamma} \int_{-x_0}^{x_0} \frac{ \gamma d 
x}{x^2 + \gamma^2}
\end{equation}
where $\gamma = \pi/2 N \lambda$. Due to the properties of the Cauchy 
distribution, one can write 
\[
\frac{1}{2}N = \frac{\lambda^{-1}}{\gamma} \int_{-\gamma}^{\gamma} \frac{ 
\gamma d x}{x^2 + \gamma^2}
\]
We can write also
\[
f_1 (x, T, N)_{T/N \to 0} \simeq \frac{2 N \lambda}{\pi} \delta (x)
\]
Thus, for small $T/N$ 
all the population tends to concentrate on the ground state,
and although not an usual Bose- Einstein condensation, we claim  that (as 
in the magnetic field case, \cite{Perez}) we have a "diffuse" condensation 
\cite{Smolenski} when $T/N$ is low enough. There is no critical point; 
there is no discontinuity in the derivative of
the specific heat in this case. {\it The condensation is the outcome of a 
continuous process in the sense that there is always some amount of particles
in the ground state, and this quantity can be increased continuously to 
reach 
macroscopically significant values, by decreasing enough the ratio $T/N$, 
i.e., even for temperatures far from zero.}
The $D = 1$ case satisfies the weak criterion at any temperature $T$, but 
not the strong one.
Closely connected with the $D = 1$ case is the problem of condensation of 
a gas of charged particles in presence of a strong magnetic field 
\cite{Perez}, where all the previous considerations apply, by 
substituting $\gamma = 2 M eB T/\hbar c^2 N$, where $e$ is the electric 
charge and $B$ the magnetic field in Gauss. If $e B \hbar/M c T \gg 1$,
the system is confined to the Landau ground state $n = 0$, and the 
problem can be treated in close connection to  $D = 1$ case. 
We will consider the model of a gas of kaons in a neutron star 
\cite{Cleymans}. By taking a density $N \simeq 10^{44}$ cm${}^{-3}$,
$T \simeq 10^8$${}^{\circ}$K and local magnetic field $B \sim 10^{14}$e G,
the condition of quantum degeneracy  are exceedingly satisfied. In 
that case
$\gamma = 10^{-30}$. By taking the dimensions of the star as $10^7$ cm,
the discrete quantum states would be spaced by an amount of $\delta p =
10^{-34}$ gcm/s. One half of the total density would be distributed in
$\eta = 2 \gamma/\Delta p = 10^4$ quantum states. The ground state density
with zero momentum $p$ along the magnetic field, can be estimated then as 
$\Delta N = \eta^{-1} N = 10^{40}$, leading to observable effects: 
superfluidity and strong diamagnetic response to the applied field.
(the magnetization ${\cal M} = e B \hbar/2 M c \simeq 10^{16}$ G would 
exceed in two orders of magnitude the microscopic applied local field and 
would be the
preponderating field). However, the quantity $\Delta N/N = \eta^{-1}=
\pi \hbar {\cal N} c/M e B T V^{4/3}$ tends to zero in the thermodynamic 
limit, and some of the physical effects or condensation would appear 
softed or even erased in that limit.

Let us now turn our attention to the $D = 2$ case. Here $l = N^{-1/2}$
The distribution is 
\[
f_2 (x, T, \bar \mu) = \frac{x}{e^{x^2 + \bar \mu} - 1}
\]
and has always an absolute minimum at $x = 0$, (the density vanish 
in the ground state) and a maximum at a value of $x > 0$ being a solution 
of $ 
e^{x^2 + \bar \mu} = \frac{1}{1 - 2 x^2}$. We can write the density as 
\begin{equation}
N = 2 \lambda^{- 2} \int_0^{\infty} \frac{e^{-(x^2 + \bar \mu)} x dx}
{ 1 -  e^{-(x^2 + \bar \mu)}} = 2 \lambda^{-2} \ln (1 - e^{- \mu}). \label{2}
\end{equation}
We have $\bar \mu = - \ln (1 - e^{- N \lambda^2/2})$; thus
as $T/N \to 0$, $\bar \mu \to e^{-N h/4 \pi M T}$. The solution $\bar 
\mu = 0$  makes the density $N$ in (\ref{2}) to
diverge, as in the $D = 1$ case, except for $T = 0$. At nonzero $\bar \mu$, 
the population in a closed small neighborhood of $x = 0$ is strictly
zero, but the amplitude of this interval decreases as $\bar \mu \to 0$,
that is, the maximum of the density in momentum space is reached at a 
value of the momentum $x_{max} \neq 0$ which {\it decreases} as $T/N \to 0$. 
The maximum of the density is in a neighborhood but 
not strictly in the ground state. None of the weak and strong criteria 
are satisfied at any temperature $T \neq 0$. Thus, there is no strict 
Bose-Einstein condensation in the 
sense that the value of the density in the ground state is zero for any $T/N
\neq 0$. As one can write for $x$ small
\[
\lim f_2 (x, T, \bar \mu)_{x \to 0} = \lim \frac{1}{\pi} \frac{x}{x^2 + 
\bar \mu} = \delta (\sqrt{\bar \mu}),
\]
i.e., the density at $x = 0$ is nonzero only for $\bar \mu = 0$. 
But as the maximum of the density increases continuously by decreasing 
$T/N$, being located at  $x_m \to  0$, (the convergence to $x = 0$ being 
non-uniform), we have  a sort of quasi-condensate, in the spirit of the 
weak criterion, i.e. most of the 
density can be found concentrated in a small interval of values of 
momentum around the $p = 0$ state at arbitrary small temperatures.

The pressure $P$ (= force$/L$) and energy desntity in such case are
approximately given by \begin{equation}
P = 2 \lambda^{-2} T e^{- N \lambda/2}( N \lambda/2 - 1) \hspace{1cm} U = P,
\end{equation}
and obviously vanish in the limit $T \to 0$. For high $T$, these 
quantities vary linearly with $T$, as can be easily verified. The 
behavior of $C_v$ is roughly similar to the the $D = 1$ case.

\section{The case  $D > 3$ }

In the case D $>$ 3, the density in momentum space reads,
\[
f_D (x, T, \bar \mu) =\frac{x^{D - 1}}{e^{x^2 + \bar \mu} - 1}
\]
This function has an absolute minimum at $x = 0$ and an absolute maximum
at $x = x_{max} \neq 0$ given by the nonzero solution of the equation 
$e^{x^2 + \bar \mu} = 1/[1 - 2 x^2/(D - 1)]$. {\it The density is, thus, zero
at the ground state, and in a small  neighborhood of it, and in this 
sense differs from the D = 3 and the D = 1 cases}. The total density is given by

\begin{equation}
N = \frac{\lambda^{-D}}{\Gamma (D/2)} \int_0^{\infty} f_D d x = \lambda^{-D} 
g_{D/2} (z) \label{D}
\end{equation}

 In this case, 
the density $\bar 
\mu$ decreases as $T \to 0$ and $N$ converges for exactly $\bar \mu = 0$.
Thus, there is a nonzero critical temperature $T_c$
such that $\bar \mu (T_c) = 0$. Then for $T < T_c$ (\ref{D}) is unable to 
account for the total density and if conservation of particles is 
imposed, one must admit that the lacking density
$N_0 = N - N'$ is exactly at the ground state. Thus, although the density 
of particles in an (open) neighborhood of the ground state is zero, exactly
at the ground state it is given by 
$2 N [1 - (T/T_c)^{D/2}]\lambda^D \delta (x)$. 

At the critical temperature $N_c \lambda^D = \zeta(D/2)$, which tends to 
unity with increasing $D$. We may conclude that the $D$-dimensional gas 
becomes less degenerate with increasing $D$. For any dimension, the 
relation between energy and pressure $P = 2 U/D$ holds. Also, below the 
critical temperature, $C_v \sim T^{D/2}$.
We see that the $D > 3$ 
case satisfies the strong but not the weak criterion for $T < T_c$.
Our results for $D \geq 3$ are in agreement with those obtained in ref.
\cite{Kirsten}.

Expression (\ref{D}) is valid for continuous D. It can be easily checked that
for $1 > D > 0$ the quantity $f_D (x)$ diverges for $x = 0$, and $N$ 
remains finite for $\bar \mu \neq 0$. Thus the "diffuse" condensation 
takes place in the interval $1 \geq D > 0$. The $D = 2$-like behavior 
occurs for $2 \geq D > 1$, whereas, as demonstrated by May \cite{May},
usual condensation occurs for $D > 2$. However, for $2 < D < 3$, both 
criteria, weak and strong are satisfied, $f_D$ being divergent at $x = 0$,
whereas $N$ remains finite. 

\section{The relativistic case}

We shall revisit the relativistic case. Here the conservation of 
particles must reflect some invariance property of the Lagrangian.
We are keeping in mind the simplest case of a charged massive scalar field.

In that case, the conserved quantity, derived from the Noether theorem 
is the charge (in $D$ spatial dimensions) and as 
different from \cite{May}, we must include
 the contribution of antiparticles: this is a natural 
consequence of a 
relativistic finite-temperature treatment of the problem and 
we cite only some authors \cite{Fradkin}, \cite{Haber}, \cite{Kapusta}. At 
high $T$ one 
must consider the natural excitation of pairs particle-antiparticle, and 
the conserved quantities depend usually from  the difference of their
average densities.

\begin{equation}
Q = i \int_0^{\infty} j_0 (x) d^{D}x
\end{equation}

where 
\[
j_\nu =  \psi^{*} \partial_\nu\psi -  (\partial_\nu \psi^{*}) \psi
\]

After building the density matrix for the Grand Canonical ensemble, one can 
write the thermodynamic potential, and from it the conserved charge as an 
expression which contains the difference of average number of particles 
minus antiparticles:

\begin{equation}
<Q> = \frac{2 \pi^{D/2}T^D}{\Gamma (D/2) c^D \hbar^{D}}\int_0^{\infty} 
x^{D-1} dx (n_p - n_a), 
\label{RC} 
\end{equation}
where $ n_p = [(e^{E - \bar \mu} - 1]^{-1}$, 
$n_a = [e^{E + \bar \mu} - 1)^{-1}]$ are the particle and antiparticle 
densities, 
$ E = (x^2 + \bar M^2)$, and $x = p/T$, $\bar M = M c/T$. Condensation 
in $D \geq 3$ occurs for $\bar \mu = \to \bar M$. For $D < 3$ the 
condensation  is 
very well reproduced by the infrared (non-relativistic) limit already 
seen, by taking the chemical potential as $\mu' = \mu - M$.

A very interesting case occurs when $M = 0$. In that case (\ref{RC}) demands
$\mu = 0$ for not to have negative population densities of particles or 
antiparticles. All the charge must be concentrated in the condensate and 
the critical temperature for condensation, as suggested in \cite{Haber} is 
$T = \infty$. 

\section{The case  $M = M(T)$}

In some systems the interactions at high temperature behave in such a way 
that 
can be described effectively as free particle systems with variable mass; 
i.e., the 
temperature-dependent interaction leads to the arising of a mass $M = M(T)$.
If $M(T) \to 0$ we have BEC with increasing temperatures, as discussed in
\cite{Chai}. We have in that case two regions for condensation: the low 
temperature and the extremely high one. It is interesting to consider 
also the case in which $M(T)$ decreases enough to have conditions 
for condensation in some interval $T_1 \leq T \leq T_2$,
where $T_1 \neq 0$, $T_2 \neq \infty$ are the two critical points. 
For D= 3 we would have 
condensation in some "hot" interval of temperatures; i.e. superfluid or 
superconductive effects may appear in some intervals of temperature even 
far from $T = 0$. All  our previous considerations for condensation 
in $D \neq 3$, would also be valid in such case.

 The author thanks  A. Cabo and K. Kirsten for very interesting 
comments and to G. Senatore for a discussion.


\begin{thebibliography}{99} 
\frenchspacing
\bibitem{Anderson} M. H. Anderson, J. R. Ensher, M. R. Matthews, C. E. 
Wieman, E. A. Cornell, {\it Science} {\bf 269} (1995), 198.
\bibitem{Pathria} R.K. Pathria, {\it Statistical Mechanics, Pergamon Press},
Oxford, (1972).
\bibitem{Landau} L. D. Landau, E. M. Lifshitz, {\it Statistical Physics, 
Pergamon Press}, Oxford, (1980)
\bibitem{Haber} H. E. Haber and  H.A. Weldon, {\it Phys. Rev. Lett.}{\bf 
46} (1981), 1497;  {\it Phys. Rev.}{\bf D 25} (1982), 502.
\bibitem{Kapusta} J. Kapusta, {\it Phys. Rev.}{\bf D 24}, (1981), 426.
\bibitem{Wagner} N.D. Mermin and  H. Wagner, {\it Phys. Rev. Lett}{\bf 17}
(1966), 1133.
\bibitem{Uhlenbeck} M. Ziff, G. E. Uhlenbeck and M. Kac, {\it Phys. 
Rep.}{\bf 32} (1977), 169.
\bibitem{Coleman} S. Coleman {\it Commun. math. Phys.}{\bf 31} (1973) 259.
\bibitem{May} R. M. May, {\it Phys. Rev.} {\bf 135} (1964) A1515.
\bibitem{Kirsten} K. Kirsten and D. J. Toms {\it Phys. Lett.}{\bf B 368} 
(1996) 119.
\bibitem{Perez} H. Perez Rojas, {\it Phys. Lett.}{\bf B}(1996) (to be pub.)
\bibitem{Smolenski} G. A. Smolenski and V. A. Isupov, {\it Sov. Journal of 
Techn. Phys.} {\bf 24} (1954) 1375; R. L. Moreira and R. P. S. M. Lobo,
\bibitem{Cleymans} J. Cleymans and D. W. Oertzen, {\it Phys. Lett.}{\bf B
 249} (1990), 511.
{\it Jour. Phys. Soc. Japan}{\bf 61} (1992), 1992.
\bibitem{Fradkin} E. S. Fradkin, {\it Proceedings from the P. N. Lebedev 
Physical Institute N0. 29, Cons. Bureau, N.Y.} (1967).
\bibitem{Chai} H. Perez Rojas and O. K. Kalashnikov, {\it 
Nucl. 
Phys.}{\bf B 293} (1987) 241; H. Perez Rojas, {\it Phys. Lett.}{\bf A 
137}(1989), 13; M. Chaichian, R. Gonzalez Felipe, H. Perez Rojas, {\it Phys. 
Lett.}{\bf B 342} (1995) 245.
\end{thebibliography}
\end{document}